%% file: main.tex
\documentclass[sigconf]{acmart}

\usepackage{booktabs} 
\usepackage{url}
\usepackage{multirow}
\usepackage{balance}
\usepackage{soul}
\usepackage{color}
\usepackage{colortbl}

\newcommand{\urlBDCI}{\url{http://www.lta.disco.unimib.it/tools/bdci}}
\newcommand{\CHANGE}[1]{\textcolor{black}{#1}}

\setcopyright{acmlicensed}
\acmPrice{15.00}
\acmDOI{10.1145/3106237.3106296}
\acmYear{2017}
\copyrightyear{2017}
\acmISBN{978-1-4503-5105-8/17/09}
\acmConference[ESEC/FSE'17]{2017 11th Joint Meeting of the European Software Engineering Conference and the ACM SIGSOFT Symposium on the Foundations of Software Engineering}{September 4--8, 2017}{Paderborn, Germany}

\begin{document}
\title{BDCI: Behavioral Driven Conflict Identification}

\author{Fabrizio Pastore, Leonardo Mariani, Daniela Micucci}
\affiliation{%
  \institution{University of Milano-Bicocca, DISCo}
  \city{Milan}
  \country{Italy}
  \postcode{20126}
}
\email{{pastore,mariani,micucci}@disco.unimib.it}

\begin{abstract}
Source Code Management (SCM) systems support software evolution by providing features, such as version control,\linebreak branching, and conflict detection. Despite the presence of these features, support to parallel software development is often limited. SCM systems can only address a subset of the conflicts that might be introduced by developers when concurrently working on multiple parallel branches. In fact, SCM systems can detect \emph{textual conflicts}, which are generated by the concurrent modification of the same program locations, but they are unable to detect \emph{higher-order} conflicts, which are generated by the concurrent modification of different program locations that generate program misbehaviors once merged. Higher-order conflicts are painful to detect and expensive to fix because they might be originated by the interference of apparently unrelated changes.

In this paper we present Behavioral Driven Conflict Identification (BDCI), a novel approach to conflict detection. BDCI moves the analysis of conflicts from the source code level to the level of \emph{program behavior} by generating and comparing behavioral models. 
The analysis based on behavioral models can reveal interfering changes as soon as they are introduced in the SCM system, even if they do not introduce any textual conflict.

To evaluate the effectiveness and the cost of the proposed approach, we developed BDCI$_f$, a specific instance of BDCI dedicated to the detection of higher-order conflicts related to the functional behavior of a program. The evidence collected by analyzing multiple versions of Git and Redis suggests that BDCI$_f$ can effectively detect higher-order conflicts and report how changes might interfere.
\end{abstract}


%


\thanks{This work has been partially supported by the H2020 Learn project, which has been funded under the ERC Consolidator Grant 2014 program (ERC Grant Agreement n. 646867).}

\maketitle

\input{Introduction}

\input{BDCI}

\input{example}

\input{modelGeneration}
\input{modelAnalysis}

\input{experiments}
\input{related}
\input{conclusions}
\input{artifact}

\bibliographystyle{ACM-Reference-Format}

\bibliography{biblio}

\end{document}

%% file: Introduction.tex


\section{Introduction}\label{sec:introduction}

Software development often requires working with multiple versions and multiple development branches in parallel~\cite{Perry:2001:PCL:383876.383878}. For example, developers may need to implement a bug fix without interrupting the development of new features. This is often done by creating a new development branch dedicated to bug fixing, which is later merged with the main branch. The use of the branching logic gives flexibility to developers, but also introduces issues. In particular, merge operations might be extremely painful due to any conflicting and interfering changes that need to be correctly managed in the merged version.
For instance, Brun et al. reported that 24\% of merge operations in open source projects generate textual conflicts, build problems, and test failures~\cite{Brun:ESECFSE:CollaborationConflicts:2011}; Kasi et al. analyzed four popular open source projects hosted on GitHub and concluded that conflicts occur regularly, with a frequency ranging from 34\% to 54\% of the merge operations~\cite{kasi:cassandra:ICSE2013}; Microsoft developers reported that most of the time dedicated to merge operations is spent on resolving conflicts and verifying correctness~\cite{Bird:FSE:ValueOfBranches:2012}. The relevance of the problem is further confirmed by the results of a recent survey conducted internally to Microsoft, which indicates that novel solutions for conflict detection and resolution are perceived as extremely useful in practice~\cite{Lo:PractitionersRelevanceSEResults:FSE:2015}. 

Source Code Management (SCM) systems support developers during their activity providing features such as version control, branching, and conflict detection~\cite{Estublier:ImpactSCM:TOSEM:2005,Grinter:ConfigurationManagement:COCS:1995,Perry:2001:PCL:383876.383878}. When two branches are merged, SCM systems can detect and report \emph{textual conflicts}, that is, changes occurring in different branches but targeting the same locations in the source files. 
Since resolving these conflicts might require significant effort, especially when done after many changes have been accumulated on the individual branches~\cite{brown:software:2002}, workspace awareness has been investigated as a possible solution~\cite{biehl:fastdash:SIGCHI:2007,Appelt:BSCW:CCTTI:1999,Estublier:Celine:WKSCM:2005,Fitzpatrick:Elvin:CSCW:2002}. The key idea is to make developers aware of the ongoing changes implemented by the other developers in their workspaces so that they can timely react in case of potential textual conflicts.


These techniques are useful but are limited to textual conflicts and cannot detect \emph{higher-order conflicts}~\cite{Brun:EarlyDetection:TSE:2013,Bang:WorkspaceAwarenessModeling:CSCW:2012}. A higher-order conflict takes place when merging two program versions in two different branches produces a faulty program without causing any textual conflict~\cite{Brun:EarlyDetection:TSE:2013,Bang:WorkspaceAwarenessModeling:CSCW:2012,Brun:ESECFSE:CollaborationConflicts:2011,Horwitz:NoninterferingVerions:TOPLAS:1989,Mariani:ConflictDetection:ISSRE:2014}. This happens when changes affecting different program locations interfere. For example, if two developers independently modify the implementation of a program for sorting the items in a collection by changing the type of elements that can be added to the collection in one case, and changing the sorting algorithm in the other case, the two changes, although targeting different areas of the code, might not integrate well. This may occur because the new version of the sorting algorithm might be unable to properly sort the new types of items that can be added to the collection. These cases are particularly hard to detect because cannot be recognized by simply considering the changed lines of code, but it is necessary to take into account the semantics of the program.

Detecting and resolving higher-order conflicts is painful and time consuming on one side~\cite{Horwitz:NoninterferingVerions:TOPLAS:1989}, but also extremely actual and important on the other, due to the increasing popularity of modern SCM systems\footnote{According to recent surveys, 62\% of the Debian projects use modern SCM environments~\cite{Zacchiroli:Debian:WEBSITE}, and 40\% of the medium and large enterprises surveyed in~\cite{Noyes:GITAdoption:WEBSITE} use Git.}, such as Git~\cite{GIT:WEBSITE}, that encourage the use of the branching logic, and thus facilitate the introduction of both textual and higher-order conflicts.

Some approaches address higher-order conflicts by enriching workspaces with the capability to report concurrent chan\-ges to dependent artifacts, such as changes to related clas\-ses~\cite{dewan:semi-synchronous:ECSCW2007,schummer:TUKAN:ECSCW2001,sarma:palantir:TSE2012,Hattori:CollaborativeSWDevelopment:ICSE:2010}. These techniques may facilitate the detection of higher-order conflicts, but require developers to \emph{manually} check all the reported changes to determine if they generate any actual higher-order conflict. Moreover, reporting to users any potential conflict, even resulting from indirect dependencies, may generate many false alarms, while reporting potential conflicts resulting only from direct dependencies may miss several higher-order conflicts.

Speculative merging has been proposed to mitigate these issues and more efficiently address higher-order conflicts~\cite{Brun:EarlyDetection:TSE:2013,Brun:ESECFSE:CollaborationConflicts:2011,Guimaraes:EarlyConflictDetection:ICSE:2012}. The underlying assumption is that interfering changes should result in test failures. Thus, to detect higher-order conflicts, the code in the developers' workspaces is merged locally with the code extracted from other parallel development branches and then locally built and tested.

\CHANGE{Speculative merging extensively uses the resources in the developers' machines for the
merge, build, and test operations and its effectiveness depends on the
fault-detection capability of the test cases. A poor or \emph{limited set of
tests and assertions} might severely limit its effectiveness. Moreover, when a test
failure is reported, developers still have to manually inspect the
concurrent changes to understand how they interfere, and this might be
expensive for non-trivial changes. 
Online continuous integration systems, such as Travis~\cite{TRAVISWEB}, provide
speculative merging as a service, thus eliminating the problems related to the consumption of
computational resources on developers' machines. 
However, except for
resource consumption, these systems are still affected by the same limitation
than speculative merging, that is, its capability to detect higher-order conflicts depends on
the number and extensiveness of the checks (e.g., the assertions) implemented in the test cases.}

In this paper we present \emph{Behavioral Driven Conflict Identification} (BDCI), a \emph{multi-branch} server-side dynamic analysis technique that \emph{detects higher-order conflicts} in parallel development bran\-ches by explicitly \emph{deriving and analyzing a representation of the program behavior}. 
The key idea is that by implementing conflict detection at the level of the program behavior, it is possible to detect interfering changes \emph{at the place where the interference can be observed}, regardless the place where the changes have been implemented. For instance, changes to apparently unrelated classes may cause interference in a same program location. This interference might be hard to detect by looking at program dependencies, but it is easy to detect when looking at the program behavior because both code changes would affect exactly the same behavior.

BDCI leverages specification mining~\cite{Ernst:Daikon:IEEETSE:2001,Lorenzoli:gkTail:ICSE:2008} to automatically generate models that represent the behavior of the program, and model analysis~\cite{Microsoft:Z3:WEBSITE} to both determine how the behavior of the program has changed along each branch and compare the modified behaviors to identify higher-order conflicts. 
The identified higher-order conflicts are returned to developers as soon as they are introduced in parallel branches, well before they produce misbehaviors in the merged version, allowing developers to timely address conflicts when they are still easy to fix.

Since BDCI compares mined models to identify higher-order conflicts, its effectiveness is not dependent on the assertions implemented in the test cases, as in speculative merging, but it simply requires the availability of some test inputs to run the program. Thus, BDCI can even work with automatically generated test cases. Moreover, BDCI provides a richer output compared to state of the art solutions. In fact, BDCI indicates how each change has affected the behavior of the program, generating information useful to quickly resolve higher-order conflicts. Finally, although this paper focuses on the functional behavior of a program, the concept of analysis defined in BDCI can be potentially exploited to investigate how changes impact other classes of behaviors, such as performance or security-related behaviors.

The empirical results obtained by analyzing multiple chan\-ges in the Git and Redis open source projects show that BDCI can reveal higher-order conflicts that cannot be detected with speculative merging. 

The main contributions of this paper are: 
\begin{itemize}
\item The definition of BDCI, a novel and general approach to the detection of higher-order conflicts.
\item The definition of BDCI$_f$, a specific instance of BDCI dedicated to the detection of higher-order conflicts on the \emph{functional} behavior of a program.
\item The release of a freely available implementation of\linebreak BDCI$_f$, which can be downloaded from \emph{\urlBDCI}.
\item An empirical validation that provides evidence of the effectiveness of BDCI$_f$.
\end{itemize}

The paper is organized as follows. Section~\ref{sec:BDCI} introduces the BDCI approach and presents BDCI$_f$.
Section~\ref{sec:example} introduces a running example. 
Section~\ref{sec:modelGeneration} describes how BDCI$_f$ generates models of the functional behavior of a program. Section~\ref{sec:modelAnalysis} presents how BDCI$_f$ uses the generated models for the detection of higher-order conflicts. Section~\ref{sec:experiments} presents the empirical results. Section~\ref{sec:related} discusses related work. Section~\ref{sec:conclusions} provides concluding remarks.

%% file: BDCI.tex
\section{BDCI}\label{sec:BDCI}

\begin{figure*}[ht]
\begin{center}
\includegraphics[width=18cm]{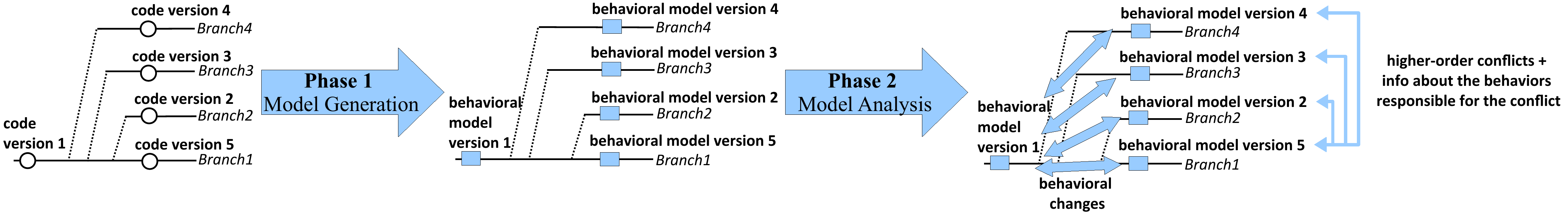}
\vspace{-0.5cm}
\caption{The BDCI approach.}
\label{fig:bdcigeneral}
\end{center}
\end{figure*}

BDCI raises conflict detection \emph{from the source code level}, as done in current SCM systems, \emph{to the behavioral level} by deriving and comparing models that represent the behavior of the software. 
The way BDCI analyzes behavioral changes is similar to the way source code changes are handled by regular SCM systems.
A commit operation may introduce some changes that modify the behavior of the software, thus causing changes in the behavioral models.
Changes introduced in parallel development branches can hence be compared according to their effect on the models. In particular, if the changes introduced in two parallel development branches affect the same behaviors in the models, the developers will likely experience an interference once the code in these branches is merged. BDCI  timely detects and signals these dangerous situations to the developers by analyzing the behavior of parallel software versions.

The BDCI approach consists of two phases, the model generation phase and the model analysis phase. These two phases are executed sequentially and might be activated according to different strategies. For instance, they can be executed \emph{after a new version is committed} to the SCM system, to detect higher-order conflicts between the committed version and the other versions in the parallel development branches, \emph{after a pull-request or a merge operation}, to detect conflicts between the changes accumulated in two specific branches (pull-requests and merges are the operations typically used to merge contributions from different developers), or \emph{every night}, to detect conflicts among the versions committed during the day and the ones in the other parallel branches.

Figure~\ref{fig:bdcigeneral} shows the BDCI approach applied to a software system developed along four different branches created from a common base version. 

In the \emph{model generation phase}, BDCI mines the models useful to detect higher-order conflicts using a specification mining solution. Depending on the aspect that is investigated, these models could target a specific class of program behaviors. For instance, models may potentially represent the functional behavior, the performance, the security-related behavior of an application, or a combination of them. BDCI generates behavioral models for all the versions that must be compared and analyzed, that is, the base version (version 1 in Figure~\ref{fig:bdcigeneral}) and the latest versions on the parallel branches (versions 2 to 5 in Figure~\ref{fig:bdcigeneral}). 


In the \emph{model analysis phase}, BDCI identifies behavioral changes and higher-order conflicts relying on the models obtained in the model generation phase. 
In particular, BDCI first identifies behavioral changes by comparing the model of the base version to the models of the versions in the parallel branches. 
BDCI then detects higher-order conflicts by identifying the parallel branches that modified the same behaviors. Figure~\ref{fig:bdcigeneral} shows the comparisons performed 
between version 5 and the versions in the other parallel branches.


Model analysis naturally detects changes that interfere according to the aspect represented in the models. If the models represent the functional, performance, or security behavior of the application, the higher-order conflicts will consequentially target the functional, performance, or security behavior of the application, respectively.

In addition to reporting the presence of higher-order conflicts, BDCI reports information about the program behaviors responsible for the conflict. Note that interfering chan\-ges may often produce observable effects in program locations that are different from the places where the code has been modified. BDCI always returns information about the higher-order conflicts by referring to the locations where they produce \emph{observable effects} that can be exploited by developers to drive the conflict resolution process. 


%

\medskip

\begin{figure*}[ht]
\begin{center}
\includegraphics[height=6.5cm]{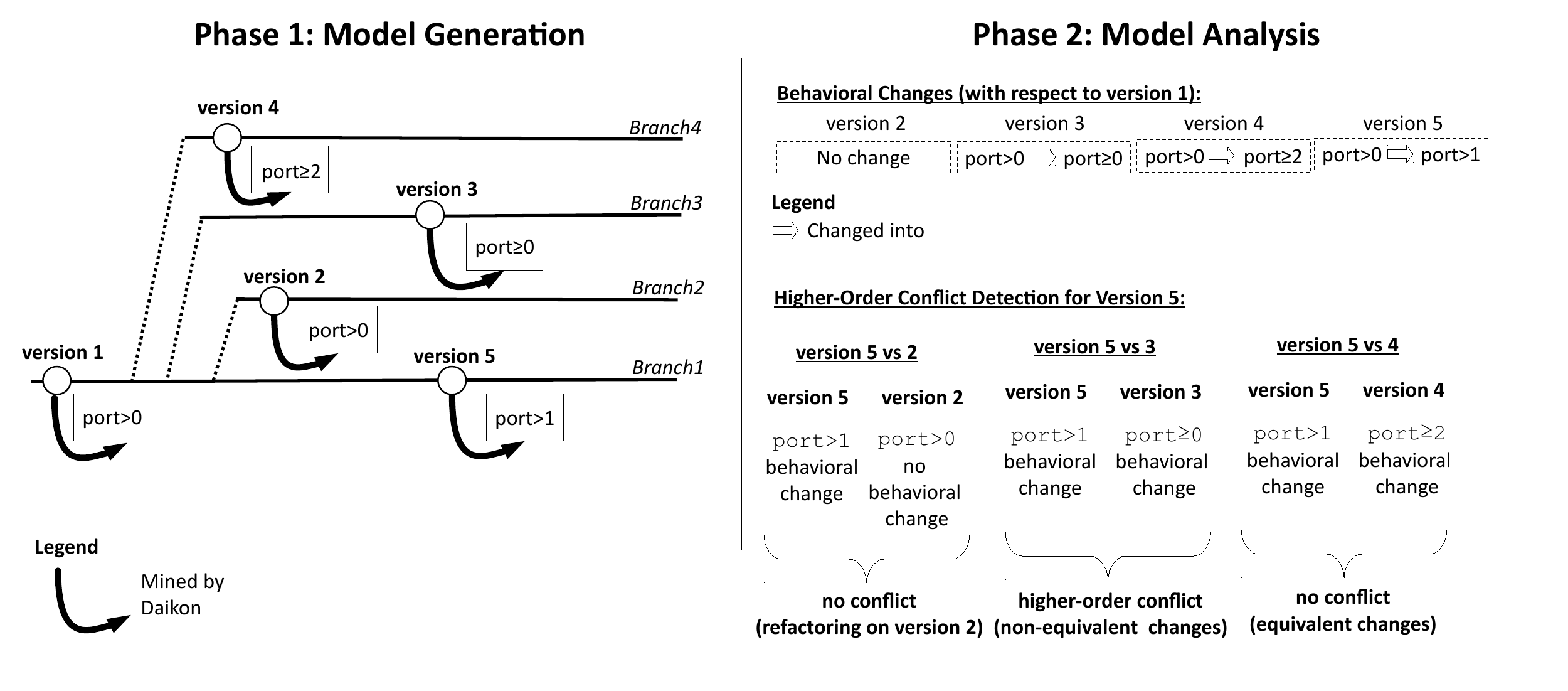}\vspace{-0.3cm}
\caption{The BDCI$_f$ approach.}
\label{fig:bdci}
\end{center}
\end{figure*}

In this paper we focus on detecting higher-order conflicts related to the \emph{functional behavior} of a program. In particular, we show how BDCI can be used to discover interfering changes that may result in annoying functional faults when merged into a same program version. This specific instance of BDCI is called BDCI$_f$. Hereafter, we will use the term BDCI when describing general aspects, and the term BDCI$_f$ when referring to the specific instance of BDCI for detecting higher-order conflicts in the functional behavior of a program. 

To capture the functional behavior of an application,\linebreak BDCI$_f$ focuses on function pre- and post-conditions. In particular, BDCI$_f$ uses the Daikon specification mining technique to discover properties that represent the values that can be assigned to function parameters~\cite{Ernst:Daikon:IEEETSE:2001}, and the Z3 theorem prover to identify changes and conflicts between sets of properties associated with different program versions~\cite{Microsoft:Z3:WEBSITE}. 
Hereafter, we will interchangeably use the terms models and properties, because program properties, and in particular pre- and post-conditions, are the kind of models derived by BDCI$_f$.

Figure~\ref{fig:bdci} graphically illustrates how BDCI$_f$ works, with specific reference to a software project with four parallel development branches all created from the same base version, version 1. We present BDCI$_f$ considering the case it is executed after a commit operation. When executed after a pull-request or overnight, BDCI$_f$ simply runs the described analysis on all the relevant program versions. We suppose version 5 is the latest version committed to the SCM system. For simplicity, we present the analysis only considering one property about the parameter \texttt{port}. 

\smallskip

In the \emph{model generation phase}, BDCI$_f$ uses Daikon to derive properties for both the version under analysis (version 5 in the example) and the latest versions available in the other parallel branches (versions 2, 3, and 4 in the example). 

BDCI$_f$ also derives properties for version 1, which is the latest common ancestor version of the compared versions. This is necessary because BDCI$_f$ reveals higher-order conflicts by comparing the program properties \emph{changed} in version 5 to the properties \emph{changed} in versions 2, 3, and 4, one at time.
The set of changes relevant to a comparison between two program versions in two different branches consists of all the changes that have been accumulated on the individual branches. The effect of these changes on the behavior of the application can be determined by comparing the properties mined for the versions under analysis to the properties mined for their latest common ancestor version. 
Since in the example all the versions are obtained from version 1, version 1 is the latest common ancestor version.


The left side of Figure~\ref{fig:bdci} shows the four branches, the five program versions, and the properties derived for parameter \texttt{port}. Since Daikon mines program properties from traces collected during tests execution, we assume that test cases are available under the SCM system. 

\smallskip

In the \emph{model analysis phase}, represented on the right side of Figure~\ref{fig:bdci}, BDCI$_f$ first determines the behavioral changes and then identifies the higher-order conflicts. To determine the behavioral changes, BDCI$_f$ compares the properties derived for each program version in the parallel branches, versions 2 to 5 in the example, to the properties derived for the latest common ancestor version, version 1 in the example. The top-right side of Figure~\ref{fig:bdci} shows the behavioral changes identified by BDCI$_f$. Only the changes in version 2 do not affect the property $\textit{port}>0$, which holds both in version 1 and version 2 \CHANGE{(this may happen for example in the case of pure refactorings)}. 

A higher-order conflict occurs when the same property is changed in two different parallel branches. To discover if version 5 introduced a higher-order conflict, BDCI$_f$ compares the behavioral changes introduced in version 5 to the behavioral changes in the other parallel branches. The right-bottom part of Figure~\ref{fig:bdci} reports the outcome of the comparison of version 5 to versions 2, 3, and 4.

The comparison between version 5 and version 2 reports no conflicts because the property has been changed in a branch only, while the parallel modification of a same behavior is necessary for the generation of a higher-order conflict. 

\CHANGE{Since version 5 and version 4 are characterized by parallel changes that affect the same property in the same way, that is, they lead to new properties that are equivalent, the comparison between these two versions does not produce any conflict. 
In our empirical experience, we observed this case several times. The most frequent reason for parallel and equivalent changes is the deployment of a same patch to a critical fault on all the active development branches.}

Finally, a higher-order conflict is reported by the comparison between version 5 and version 3. The changes in version 5 modify the values that can be assigned to parameter \texttt{port} from $\textit{port} > 0$ to $\textit{port} > 1$. While the changes in version 3 modify the values that can be assigned to parameter \texttt{port} from $\textit{port} > 0$ to $\textit{port} \geq 0$. Changes in versions 5 and 3 could thus interfere on variable \texttt{port} and generate unpredictable results once merged. BDCI automatically identifies and reports this dangerous situation.

\begin{figure*}[ht]
\begin{center}
\includegraphics[width=14cm]{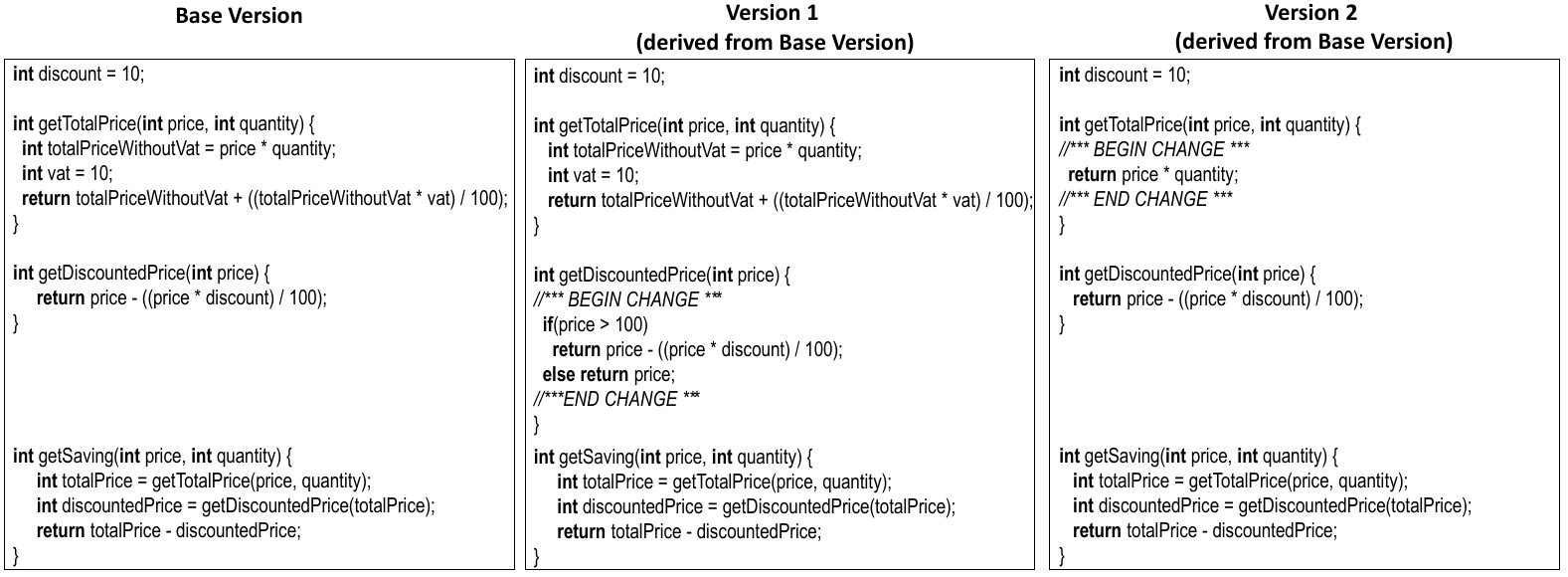}
\caption{An example of a higher-order conflict.}
\label{fig:example}
\end{center}
\end{figure*}


%% file: example.tex
\section{Running Example}\label{sec:example}

To illustrate BDCI$_f$ we use the simple example of higher-order conflict shown in Figure~\ref{fig:example}. The example consists of three program versions: the \emph{Base Version}, and two derived versions, \emph{Version 1} and \emph{Version 2}, implemented in two parallel development branches. Both Version 1 and Version 2 are derived from the Base Version, which represents their latest common ancestor version.

The Base Version is composed of three C functions. The\linebreak \texttt{getSaving} function takes a \texttt{price} per unit, a \texttt{quantity}, and a \texttt{discount} as parameters and computes the money saved thanks to the discount. Some operations are delegated to auxiliary functions. The \texttt{getTotalPrice} function computes the total price \CHANGE{(in cents)} for the considered quantity of items taxes included. The\linebreak \texttt{getDiscountedPrice} applies the discount rate to the total price.

\vfill
Version 1 implements a new policy for the discount, that is, the discount is applied only if the price of the items is higher than \CHANGE{1000 cents (i.e., 10 euros)},  taxes included. In Figure~\ref{fig:example} the changed code is the one between the comments \texttt{\small*** BEGIN CHANGE ***} and \texttt{\small*** END CHANGE ***}.

\vfill
Version 2 implements the new requirement of computing savings without considering taxes. Thus, the code that sums taxes to the total price is removed from the \texttt{getTotalPrice} function. The changed code is again indicated with \texttt{\small*** BEGIN CHANGE ***} and \texttt{\small*** END CHANGE ***}.

\vfill
Note that Version 1 and Version 2 are correct programs that properly implement the respective requirements. Also note that the two parallel changes considered in the example target different functions, and thus do not produce any textual conflict when Version 1 and Version 2 are merged together (the functions could even be in different files). However, merging Version 1 with Version 2 does not produce a correct program. The merged version would apply the discount to prices with a value greater than \CHANGE{1000} taxes excluded, while the requirement specifies that taxes must be included in the computation. Merging these two programs requires adapting the condition in the \texttt{getDiscountedPrice} function to work with prices that do not include taxes.

\vfill
A regular SCM system would merge Version 1 and Version 2 without generating any conflict. A continuous integration system running regression test cases, as well as speculative merging~\cite{Brun:EarlyDetection:TSE:2013,Brun:ESECFSE:CollaborationConflicts:2011,Guimaraes:EarlyConflictDetection:ICSE:2012}, could reveal the problem in the merged version only if the test cases (1) cover the faulty situation and (2) include proper oracles that can detect the effect of the higher-order conflict, and only after the code has been merged. In the next sections, we show how BDCI$_f$ could report this higher-order conflict to the developers as soon as Version 1 and Version 2 have been created and before they are merged. Compared to speculative merging, BDCI$_f$ only requires the existence of test cases covering the change without requiring any oracle in the test case. For this reason, contrarily to speculative merging, if test cases that cover the changes are not present in the program test suite, BDCI$_f$ can straightforwardly work with automatically generated test cases.

%% file: modelGeneration.tex
\section{Model Generation}\label{sec:modelGeneration}

In the model generation phase, BDCI generates behavioral models for all the program versions that must be compared during the analysis phase. Depending on the policy used to execute BDCI, the set of program versions that must be compared might slightly change. 
For instance, when executed after a pull-request, BDCI compares the latest program versions available in the branches that must be merged, and when executed overnight, BDCI compares the latest program versions available in all the active branches. 

In every case, BDCI compares pairs of versions. For this reason, this section and the next section present BDCI$_f$ by describing how it compares Version 1 to Version 2 for the example program shown in Figure~\ref{fig:example}. 
The general case where multiple versions are compared can be obtained by iteratively comparing pairs of versions.



BDCI$_f$ uses Daikon as model generator~\cite{Ernst:Daikon:IEEETSE:2001}. Daikon generates Boolean expressions that represent the values that can be assigned to program variables. It takes a set of variable values observed at a given program point as input, and outputs a set of program properties that hold at the same program point. For instance, if the \texttt{getTotalPrice} function is executed multiple times with a positive \texttt{price} and a positive \texttt{quantity}, Daikon can mine that its return value is always positive. In order to obtain the samples necessary to run Daikon, BDCI$_f$ assumes test cases are available under the SCM system, which is a standard practice in version control, especially when a SCM system is equipped with continuous integration capabilities~\cite{Meyer:ContinuousIntegration:IEEESW:2014,TRAVISWEB}. In the model generation phase, BDCI$_f$ builds the system, runs the test cases, collects the values assigned to program variables at several program points, and invokes Daikon to obtain the program properties. 

Generating properties for every program location in the code requires collecting and processing a huge amount of data, and this is not feasible for non-trivial programs. To restrict the focus of the analysis, BDCI$_f$ exploits the knowledge of the changes implemented in each branch. In particular, it automatically determines the functions that have been modified by running the Unix diff program~\cite{LinuxDiff:WEBSITE} 
against the compared versions and their latest common ancestor version.


As monitoring points, BDCI$_f$ selects all the functions that might be directly influenced by the identified changes. More precisely, BDCI$_f$ collects data from the entry points and the exit points of all the functions that include modified statements, the callers and the callees of these functions. This simple criterion can be computed efficiently, guarantees the selection of a reasonably small set of program points, and focuses the analysis on the program locations that are likely to be affected by the changes. In principle, more sophisticated techniques could be used to select monitoring points, such as program slicing techniques~\cite{Weiser:Slicing:IEEETSE:1984}. However, we preferred to design a technique that is simple, fast, and returns a small set of observation points, rather than a technique that might return a large list of program locations that might make the analysis too expensive or even infeasible.

BDCI$_f$ considers entry and exit points to focus on function pre- and post-conditions, which are expected to have a quite stable behavior across versions, and ignores the body of functions where extensive changes might easily occur. The functions with a changed signature are not compared directly, but are compared using the callers and callees of these functions as observation points.

In the running example, BDCI$_f$ collects data about the\linebreak \texttt{getDiscountedPrice}, because its body has been modified in Version 1, the \texttt{getTotalPrice} function, because its body has been modified in Version 2, and the  \texttt{getSaving} function, because it invokes the modified functions. The same three functions are monitored in the Base Version, Version 1, and Version 2.

BDCI$_f$ builds all program versions relevant to the analysis (in the example, the Base Version, Version 1, and Version 2) and runs the tests available under SCM to collect data. Data collection is performed by intercepting the execution at the entry and exit points of the selected functions and logging the values of all the variables in the scope of that locations.  

Finally, BDCI$_f$ runs Daikon on the collected data to derive the pre- and post-conditions. Table~\ref{table:getSaving} reports the post-conditions that BDCI$_f$ can derive for the three monitored functions. The post-conditions in Version 1 and Version 2 that differ from the Base Version are reported in boldface with gray background. We omit pre-conditions because they are not relevant to the discussion of the running example. The label \texttt{ret} indicates a function return value.

\begin{table}[htbp]
\scriptsize
\centering
\caption{Post-conditions of the monitored functions}
\label{table:getSaving}
\begin{tabular}{l l l l}


\hline
Version
& \texttt{getTotalPrice}
& \texttt{getDiscountedPrice}
& \texttt{getSaving}
\\ \hline \hline

Base Version & $\textit{ret} > \textit{price}$ & $\textit{ret} <\textit{price}$ & $\textit{ret} > 10$ \\

Version 1 & $\textit{ret} > \textit{price}$ & \cellcolor[gray]{0.9}\textbf{$\textit{ret}  \leq \textit{price}$} & \cellcolor[gray]{0.9}\textbf{$\textit{ret} \geq 0$} \\

Version 2 & \cellcolor[gray]{0.9}\textbf{$\textit{ret} \geq \textit{price}$} & $\textit{ret} <\textit{price}$ & \cellcolor[gray]{0.9}\textbf{$\textit{ret} \geq 10$} \\
\hline
\end{tabular}
\end{table}

%% file: modelAnalysis.tex
\section{Model Analysis}\label{sec:modelAnalysis}

The model analysis phase identifies higher-order conflicts by (1) identifying the \emph{behavioral changes} in the versions that must be compared (Version 1 and Version 2 in the running example), and then (2) determining if the compared versions have \emph{changed the same behaviors}. 

To determine the behavioral changes, BDCI compares the models derived for the versions under analysis against the model derived for their latest common ancestor version. In the running example, BDCI$_f$ checks equivalence between the pre- and post-conditions derived for the \texttt{getTotalPrice}, the \texttt{getDiscountedPrice}, and the \texttt{getSaving} functions in Version 1 and 2, and the pre- and post-conditions derived for the same functions in the Base Version. 

According to the post-conditions reported in Table~\ref{table:getSaving}, the chan\-ges in Version 1 affect the post-conditions of the \texttt{getDiscountedPrice} and the \texttt{getSaving} functions. While the chan\-ges in Version 2 affect the post-conditions of the \texttt{getTotalPrice} and the \texttt{getSaving} functions.

\smallskip

%

To determine if changes generate any high\-er-order conflict, BDCI$_f$ identifies the pre- and post-con\-di\-tions changed in both versions. We call the set of these pre- and post-conditions the \emph{interfering region}. 
\CHANGE{A non-empty interfering region may indicate the presence of a higher-order conflict.}
In the running example, the interfering region for Version 1 and Version 2 is not empty and consists of the post-condition of function \texttt{getSaving}. 

\CHANGE{BDCI$_f$ reports a higher-order conflict only in the case the two versions behave differently, that is, when the two versions both differ from the base version and differ from each other. 
To determine if the \emph{interfering region} indicates that the two software versions behave differently, BDCI$_f$ checks the equivalence of the pre- and post-conditions in the interfering region.  In the running example, the post-conditions of function \texttt{getSaving} in Versions 1 and 2 are not equivalent, and BDCI$_f$ returns a higher-order conflict.}


Note that the knowledge of the changed properties represents a useful information to fix the higher-order conflict. In this case, the changed properties clearly show that: (1) the changes in Version 1 affect the computation of the discount (see the changed post-condition of function \texttt{getDiscountedPrice} in Table~\ref{table:getSaving}), (2) the changes in Version 2 affect the computation of the total cost (see the changed post-condition of function \texttt{getTotalCost} in Table~\ref{table:getSaving}), and (3) these two changes interfere with the computation of the saving (see the two changed post-conditions of function \texttt{getSaving} in Table~\ref{table:getSaving}). This is a valuable information for developers, who can inspect the code \emph{directly pointing at the functions that originate the conflict with the knowledge of how and where the changes interfere}.



\CHANGE{Finally, since higher-order conflicts can be identified as soon as the two versions appear in the SCM system, before they are merged together, developers can timely repair problems thus preventing the risk of experiencing failures with the merged code where the effect of multiple changes can significantly complicate failure analysis.} 


%% file: experiments.tex
\section{Empirical Evaluation}\label{sec:experiments}

In this section, we evaluate the \emph{effectiveness} and the \emph{cost} of BDCI$_f$. We investigate the effectiveness by comparing BDCI$_f$ to state of the art me\-thods based on speculatively merging~\cite{brun:crystal:ESECFSE2011,Brun:EarlyDetection:TSE:2013,Guimaraes:EarlyConflictDetection:ICSE:2012,guimaraes:towards:ICSE_CHASE2010}. In particular, we performed two studies. The objective of the first study is to investigate if BDCI$_f$ can reveal real higher-order conflicts that cannot be detected with speculative merging. To this end, we analyzed 35 parallel changes in the Git~\cite{GIT:WEBSITE} version history, and 10 parallel changes in the Redis~\cite{REDISWEB} version history, for a total of 45 parallel changes analyzed. We found $7$ higher-order conflicts in Git and 1 higher-order conflict in Redis that could not be revealed with speculative merging due to the limited capability of the test cases to detect behavioral changes.

To compare BDCI$_f$ and speculative merging not only on the revealed higher-order conflicts but also on the \emph{missed} higher-order conflicts, we performed a second study. Since we do not know in advance all the higher-order conflicts present in third-party software, we used conflicts \emph{injected} in Git for this study. Results show that the two approaches are complemental.

Finally, we investigate cost by reporting data about the runtime cost of the technique.




In the following, we briefly describe our prototype implementation, we report the studies with injected and real higher-order conflicts, and we conclude discussing threats to validity. 

\subsection{Prototype Implementation}
Our prototype implementation targets C programs and consists of two main components, the model generation component and the model analysis component. The model generation component identifies the functions that must be monitored for a given set of changes under analysis, implements the monitoring infrastructure for collecting runtime data on top of both the RADAR framework~\cite{Pastore-RADAR-2012,Pastore-RADARtoolDemo-ICSE-2013} and the PIN tool~\cite{Intel:PIN:WEBSITE}, and integrates Daikon~\cite{Ernst:Daikon:IEEETSE:2001} to generate functions pre- and post-conditions.

The model analysis component implements the identification of changed pre- and post-conditions, and integrates the Z3 theorem prover~\cite{Microsoft:Z3:WEBSITE} to check equivalence between properties. Our prototype implementation and the rest of the experimental material can be downloaded from \emph{\urlBDCI}.

\subsection{Real Conflicts} \label{sec:realCases}
To investigate how BDCI$_f$ performs with real cases we analyzed several parallel changes from the Git~\cite{GIT:WEBSITE} and Redis~\cite{REDISWEB} version history. For our experiments we selected merges of non trivial complexity (the changes in a branch must be distributed among at least 10 different files) that do not include a major redesign of the system (we limit to 500 the maximum number of functions modified by the changes in a branch). For the Git case study we selected all the merges with the above characteristics merged after Sept. 2013, for the Redis case study we selected merges pushed to the repository after Feb 2015.
To avoid the analysis of unrelated changes, we focused on changes that either target dependent code (changes of statements that use and define the same variables) or are followed by a fix commit. Some of the selected cases produced either textual conflicts, which could be detected with standard applications for version control, or test failures, which could be detected with speculative merging. In addition to these cases, the selected portion of the version history of the analyzed applications included a total of 35 cases for Git and 10 cases for Redis, 
which might potentially produce higher-order conflicts not detected with state of the art techniques. We used these cases as benchmark for BDCI$_f$ to evaluate its capability to detect higher-order conflicts that could not be detected with speculative merging.


\begin{table}[ht]
\begin{center}
\caption{Results with real higher-order conflicts} \label{tab:realConflicts}
\scriptsize
\begin{tabular}{|l |c | c | c | c|}
\hline
Application & Parallel Changes & HOC Returned & Spurious HOC & Actual HOC \\ 
& Analyzed & & & \\
\hline
Git & 35 & 9 & 2 & 7 \\
Redis & 10 & 1 & 0 & 1 \\
\hline
Total & 45 & 10 & 2 & 8 \\
\hline

\end{tabular}
\end{center}
\end{table}

Note that every selected case includes a large number of pervasive changes that must be analyzed with BDCI$_f$. In fact, the average number of functions that have been changed in two branches under comparison is 162 with a maximum of 489 for Git, and 114 changed functions with a maximum of 301 for Redis. Although the pervasiveness of the changes represents a challenging aspect for techniques that compare program versions, BDCI$_f$ coped well with this magnitude of changes, as suggested by the results that have been obtained.  

\CHANGE{We analyzed these 45 cases by running BDCI$_f$ with the test suites provided with the programs and we discovered a total of 10 higher-order conflicts that could not be detected by speculative merging, as shown in Table~\ref{tab:realConflicts}.} 
Out of the 10 higher-order conflicts, 2 conflicts were spurious anomalies. These conflicts could be quickly discarded because caused by variables that can be legally assigned with conflicting values in two distinct development branches. In our case, the higher order conflict was caused by a file descriptor variable that legally changed its value in two parallel branches. More interestingly, we discovered 8 problematic higher-order conflicts, 7 in Git and 1 in Redis. To confirm the presence of a problem in these 8 cases, we implemented a test case that covers the change and passes when executed in the versions before the merge, but fails after the merge operation. We thus confirmed that BDCI$_f$ has been able to discover 8 unknown higher-order conflicts out of the 45 merge operations that have been analyzed. We also manually analyzed the rest of the changes without finding any additional higher-order conflict, further confirming the accuracy of BDCI$_f$.

\CHANGE{To give evidence of the \emph{qualitative} effectiveness of BDCI$_f$, here we briefly discuss the 8 higher-order conflicts that have been identified.}

\CHANGE{Five higher-order conflicts affect the functionality that prints the content of the packets processed by Git to the standard output. The changes in the two branches both affect the heading printed by the function \texttt{packet\_trace}. The merged version masks the effect of the changes performed in one of the branches. BDCI$_f$ detected this interference from the pre-condition of function \texttt{trace\_strbuf\_fl}, which is used in \texttt{packet\_trace}.} A similar problem affects function \texttt{rdbCheckThenExit} that is used by Redis to trace errors in the Redis database management system.

\CHANGE{Out of the two remaining higher-order conflicts, one affects function \texttt{get\_wcwidth}, which computes the length of a Unicode character. The changes in the compared branches affect this function, producing an interference in the merged code. BDCI$_f$ could detect this problem from the pre-con\-di\-tion of function \texttt{bisearch}, which is used in \texttt{get\_wcwidth}}.
The last higher-order conflict affects functions \texttt{cmd\_clone} and \texttt{checkout} that present different post-conditions in the three versions of the software. 
This behaviour depends on changes in function \texttt{wait\_or\_whine}.

\input{tableResults}





\subsection{Injected Conflicts} \label{sec:injectedConflicts}
To generate higher-order conflicts in a controlled way, we started from two non-conflicting changes targeting depending code blocks and located in two parallel bran\-ches in Git\footnote{For reference, the date and time of the Git versions that we used are 11/1/2006 13:57 and 11/1/2006 13:41.}, and we systematically mutated one of these changes (i.e., the set of statements that can be mutated is restricted to the statements changed in a branch) using the Milu mutation testing tool~\cite{Jia:Milu:TAICPART:2008}. Since mutations always alter the statements already modified by a change, the mutated program includes the same logical change than the original program, but with a slightly modified semantics induced by the mutation. Of course, not all the mutants result in a higher-order conflict. Sometimes the mutation does not interfere with the changes implemented by the version in the other parallel branch. We manually analyzed the resulting mutants and after discarding equivalent, redundant, and mutants that do not produce higher-order conflicts, we ended up with a total of 19 cases, each consisting of two program versions that include parallel changes that produce a higher-order conflict.

We analyzed all the cases using both BDCI$_f$ and speculative merging. For BDCI$_f$, we executed the analysis and collected data about the generated properties and the detected higher-order conflicts. For speculative merging, we merged the two program versions, built the code, executed the test cases, and used test failures as detectors of higher-order conflicts. Since no test case in the Git test suite was designed to cover the changed statements, we used a set of manually designed test cases that cover the changes in the analyzed branches to run the program. 

\CHANGE{Table~\ref{tab:results} shows the results that we obtained.} Column \emph{Analyzed Case} lists the cases that have been analyzed. The first row represents the base case with no mutation, and thus with no higher-order conflicts. The successive rows list the 19 analyzed cases. Between brackets we report the mutant operator that has generated the case under analysis (see Agrawal et al. for a quick reference about mutant acronyms~\cite{Richard:designof:1989}).

Column \emph{Generated Properties} reports statistics about the number of pre- and post-conditions that BDCI$_f$ has generated for the \emph{Base Version}, 
and the versions in the two branches, distinguished as \emph{Version in Branch 1} and \emph{Version in Branch 2}. For the two compared versions, we also report the number of unchanged (\emph{same}), dropped (\emph{del}), and newly (\emph{new}) identified properties between brackets. Note that these columns report data about the number of individual properties that compose pre- and post-conditions.

Results show that BDCI$_f$ has been able to derive hundreds of properties about the values that can be assigned to program variables, intuitively suggesting that the mined models well cover the behavior of the application for the selected functions. The effect of the changes on these properties varies a lot, depending on the nature of the change. The changes for the Version in Branch 1 significantly modify the values that can be assigned to program variables, resulting in several properties that are dropped and newly mined. The changes for the Version in Branch 2 mostly alter the execution flow, with little effect on function parameters, thus only few properties are affected by the changes. In particular, no property changes in cases from Case 1 to Case 10, and only few properties change in cases from Case 11 to Case 19.

Column \emph{Changed pre/post-conditions} indicates the number of pre- and post-conditions that have been changed in each version. It is possible to notice that BDCI$_f$ has been always able to generate a small but useful number of pre- and post-conditions that suitably captured the semantics of the change for the Version in Branch 1, while it succeeded partially with the Version in Branch 2. As already mentioned, this is due to the nature of the change that mostly affects the control flow, while the properties used by BDCI$_f$ to discover higher-order conflicts mostly capture the data-flow of a program.

Finally, column \emph{Outcome} reports results about the higher-order conflicts returned by BDCI$_f$ and speculative merging. In the case of BDCI$_f$, we also report the number of higher-order conflicts returned by the technique, which intuitively represents the amount of information available to developers to identify the source of the conflict.
The last row indicates the total number of higher-order conflicts detected by BDCI$_f$ and speculative merging.

Both BDCI$_f$ and test execution had similar effectiveness revealing 9 higher-order conflicts each. However, the set of revealed conflicts is complemental,  each technique revealed 8 higher-order conflicts that the other technique has not been able to reveal and only 1 higher-order conflict has been revealed by both approaches. Two higher-order conflicts have not been detected by any technique due to the lack of test cases covering the behaviors that produce the higher-order conflicts.

The complementarity between BDCI$_f$ and speculative merging is due to the intrinsic characteristics of these approaches. When the interference in the behavior of the application produced by the higher-order conflict alters the execution flow and propagates to the output of the application, such as for cases from Case 1 to Case 10, the conflict can be easily detected by the checks performed by the test cases on the output values produced by the tested program, but it might be harder to reveal with BDCI$_f$. This happens because, due to the altered execution flow, some functions are not executed and some properties are not generated anymore, thus reducing the set of pre- and post-conditions that can be compared. In a sense, changes that significantly alter the execution flow might leave BDCI$_f$ with too few pre- and post-conditions that can be compared. Although BDCI$_f$ has not revealed these cases, they can still be potentially addressed with BDCI, for instance by deriving models that capture the sequences of operations that are executed by a program in addition to the values that can be assigned to program variables. Incorporating this kind of information in the models generated by BDCI$_f$ is part of our future work.

When the higher-order conflict is tricky to detect, such as for cases from Case 11 to Case 19, test cases may fail to reveal it. For instance, tests may fail to propagate the wrong behavior to the output, or they may check output variables that are not affected by the interference, as happen for cases from 11 to 19. On the contrary, BDCI$_f$ detects interfering behaviors using behavioral models, so it is usually enough to execute the changes to obtain models that capture their semantics, without any need of propagating the effect of the interference to the output.
In a nutshell, BDCI$_f$ does not need test cases, but only test inputs, that is BDCI needs the runtime data to generate behavioral models, but it never exploits the oracles in the test cases. Thus, BDCI might be applied to systems and components that are not maintained together with the test cases, as long as test inputs can be generated automatically. Contrarily, using test cases to exploit higher order conflicts strongly depends on the effectiveness of the test cases.

Overall, \emph{the complementarity between BDCI$_f$ and speculative merging might be exploited to effectively address most of the higher-order conflicts}. In fact, the union of the two approaches, which simply corresponds to both running the test cases and analyzing the behavior of the program with BDCI$_f$, reveals 17 out of the 19 (89\%) higher-order conflicts considered in the evaluation.

\begin{figure}[tb]
\begin{center}
\includegraphics[height=3cm]{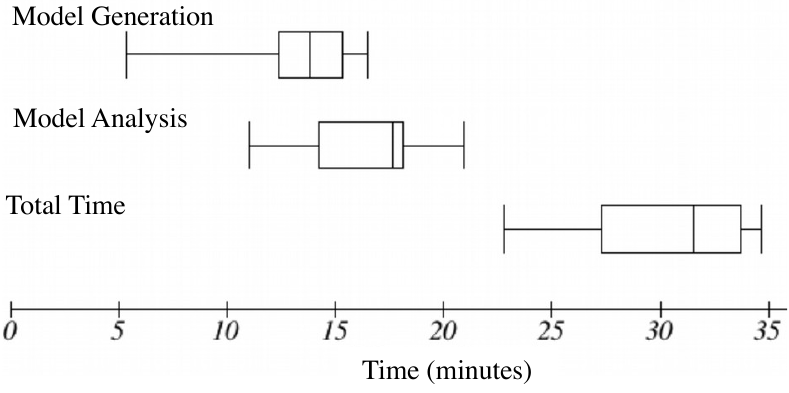}
\caption{Runtime cost of BDCI$_f$ (horizontal box-plot).}
\label{fig:cost}
\end{center}
\vspace{-0.5cm}
\end{figure}


\subsection{Efficiency} \label{sec:cost}
We executed the experiments on a machine equipped with an Intel(R) Xeon(R) CPU @ 2.53GHz. We measured the runtime cost of model generation and model analysis, as well as the total runtime cost. Figure~\ref{fig:cost} shows a box-plot of the runtime cost.

In general, model generation requires less, although comparable, time than model analysis. The major costs affecting model generation are the execution of the test cases, the monitoring overhead, and the many calls to Daikon. Model analysis requires several calls to Z3 to compare the pre- and post-conditions, to both identify behavioral changes and higher-order conflicts.

The overall runtime cost of BDCI$_f$ has ranged between 22 and 35 minutes, with a mean value of 31 minutes. These results suggest that BDCI$_f$ can be regularly applied to test overnight sessions, to pull-requests, and to sporadically analyze commit operations. 

\subsection{Threats to Validity}
The main threats to internal validity are about the correctness of our prototype implementation. 
We are confident on the correctness of our BDCI$_f$ implementation because we executed the tool on several sample cases first, to eliminate obvious faults. In addition, for all the results reported in the paper, we manually analyzed the considered case and the output produced by BDCI$_f$, thus reducing the probability that incorrect results are reported.


The main threats to external validity are about the generalizability of our findings. We experienced BDCI$_f$ with a number of parallel changes in non-trivial programs and used both real and injected conflicts for the evaluation. The injected conflicts have been obtained by modifying real changes in parallel branches, which should mitigate the risk of generating unrealistic cases. Although additional evidence should be obtained to fully generalize the results, the empirical evidence obtained so far already suggests that BDCI$_f$ could be a useful solution to improve the capability of detecting higher-order conflicts early in the development process.

%% file: tableResults.tex
\begin{table*}[ht]
\begin{center}
\caption{Results with injected higher-order conflicts} \label{tab:results}
\scriptsize
\begin{tabular}{|l || c | c | c||     c | c||     c | c| c| }

\hline

\multirow{3}{*}{\textbf{Analyzed Case}} &
\multicolumn{3}{c||}{\textbf{Generated Properties}}&
\multicolumn{2}{c||}{\textbf{Changed}}&
\multicolumn{3}{c|}{\textbf{Outcome}}\\

  &
\emph{Base} & \emph{Version in} & \emph{Version in}  &
\multicolumn{2}{c||}{\textbf{pre/post-conditions}}&
\multicolumn{2}{c|}{\textbf{BDCI$_f$}} & \textbf{Spec. Merge}\\

 &
\emph{Version} &  \emph{Branch 1} &  \emph{Branch 2} &
\emph{Version in} & \emph{Version in} &
\emph{Number of} & \emph{HOC} & \emph{HOC}\\

 &
&  \emph{(same/del/new)} &  \emph{(same/del/new)} &
\emph{Branch 1} & \emph{Branch 2} &
\emph{Conflicts} & \emph{Detected} & \emph{Detected}\\

\hline

Base case  &
156 & 158 (114/42/44) & 156 (156/0/0)&
13 & 0 &
0 & - & -\\

Case 1 (SSDL) &
11 & 8 (7/4/1) & 11 (11/0/0) &
1 & 0 &
0& NO & \textbf{YES}\\

Case 2 (OCNG) &
29 & 16 (6/23/10) & 29 (29/0/0) &
6 & 0&
0 & NO & \textbf{YES}\\

Case 3 (CRCR) &
156 & 157 (110/46/47) & 156 (156/0/0) &
14 & 0 &
0 & NO & NO \\

Case 4 (CRCR) &
29 & 17 (9/20/8) & 29 (29/0/0) &
6 & 0 &
0 & NO & \textbf{YES}\\

Case 5 (CRCR) &
29 & 17 (9/20/8) & 29 (29/0/0) &
6 & 0 &
0 & NO & \textbf{YES}\\

Case 6 (SSDL) &
137 & 126 (79/58/47) & 137 (137/0/0) &
12 & 0&
0 & NO& \textbf{YES}\\

Case 7 (OCNG) &
29 & 26 (24/5/2) & 29 (29/0/0) &
4 & 0&
0 & NO& \textbf{YES}\\

Case 8 (CRCR) &
156 & 157 (114/42/43) & 156 (156/0/0) &
13 & 0 &
0 & NO& \textbf{YES}\\

Case 9 (CRCR) &
156 & 157 (114/42/43) & 156 (156/0/0) &
13 & 0&
0 & NO& \textbf{YES}\\

Case 10 (CRCR) &
156 & 157 (114/42/43) & 156 (156/0/0) &
13 & 0 &
0 & NO& NO\\

Case 11 (SSDL) &
144 & 142 (107/37/35) & 145 (138/6/7) &
13 & 1 &
1 & \textbf{YES}  & NO\\

Case 12 (CRCR) &
156 & 157 (114/42/43) & 158 (155/1/3)&
13 & 2&
2& \textbf{YES} & NO\\

Case 13 (CRCR) &
156 & 157 (114/42/43) & 156 (155/1/1)&
13 & 2&
2& \textbf{YES} & NO\\

Case 14 (CRCR) &
156 & 157 (114/42/43) & 158 (155/1/3)&
13 & 2&
2& \textbf{YES} & NO\\

Case 15 (CRCR) &
156 & 157 (114/42/43) & 158 (155/1/3)&
13 & 2&
2& \textbf{YES} & NO\\

Case 16 (CRCR) &
156 & 157 (114/42/43) & 156 (155/1/1)&
13 & 2&
2& \textbf{YES} & NO\\

Case 17 (CRCR) &
156 & 157 (114/42/43) & 156 (155/1/1)&
13 & 2&
2& \textbf{YES} & NO\\

Case 18 (CRCR) &
156 & 158 (114/42/44) & 158 (155/1/3)&
13 & 2&
2& \textbf{YES} & NO\\

Case 19 (SSDL) &
156 & 158 (114/42/44) & 137 (129/27/8)&
13 & 4&
3& \textbf{YES} & \textbf{YES}\\

\hline

 &
 &  & &
& &
\textbf{Total} & \textbf{9} & \textbf{9}\\

\hline

\end{tabular}
\end{center}
\end{table*}

%% file: related.tex
\section{Related Work}\label{sec:related}

The problem of coordinating multiple developers and multiple teams working on a same project is well-known in software engineering. Coordination issues cover many abstraction levels, from the provisioning of basic functionalities, such as computer-mediated communication, to the provisioning of integrated operations, such as continuous coordination~\cite{sarma:categorizing:COMPUTER2010}. In this paper we focus on the problem of revealing the code-level conflicts that may result from parallel developers activity.

This problem has been already addressed with two main classes of solutions: techniques for detecting \emph{textual} (or \emph{direct}) conflicts and techniques for detecting \emph{higher-order} (or \emph{indirect}) conflicts. Techniques for detecting higher-order conflicts can be further distinguished in techniques using \emph{structural} information and techniques using \emph{behavioral} information.



\textbf{Detection of Textual Conflicts.} The ability to detect textual conflicts is a fundamental feature of every SCM system~\cite{Estublier:ImpactSCM:TOSEM:2005,Grinter:ConfigurationManagement:COCS:1995,Perry:2001:PCL:383876.383878}, such as SVN~\cite{Apache:SVN:WEBSITE} and Git~\cite{GIT:WEBSITE}. These SCM systems, although popular, are not only limited to textual conflicts but can only detect conflicts reactively, that is once two versions have been merged.

Since resolving conflicts as soon as they have been introduced in the software is definitely easier than fixing them late once software versions have been merged~\cite{brown:software:2002}, several techniques for proactive detection of textual conflicts have been defined~\cite{biehl:fastdash:SIGCHI:2007,Appelt:BSCW:CCTTI:1999,Estublier:Celine:WKSCM:2005,Fitzpatrick:Elvin:CSCW:2002}. The key idea is that conflict detection should be executed early before the code is merged, for instance every time a new version is committed to the SCM system or even when developers save changes locally to their workspaces. The result of the analysis can be made available directly in the developer workspace, to make developers aware of the ongoing activities and possible conflicts~\cite{gutwin:workspace:CHI1996}.

Some well-known techniques for proactive-detection of textual conflicts are FASTDash~\cite{biehl:fastdash:SIGCHI:2007}, which can prevent potential conflicting situations by providing a visual presentation of the developers' activities on shared files; BSCW~\cite{Appelt:BSCW:CCTTI:1999}, which provides a web-based shared workspace that integrates versioning facilities; Celine~\cite{Estublier:Celine:WKSCM:2005}, which uses a hierarchical workspace to scale to very large projects; and Elvin~\cite{Fitzpatrick:Elvin:CSCW:2002}, which provides awareness of direct conflicts.

BDCI is also a proactive conflict detection technique that can detect conflicts as soon as they are introduced in the SCM system. Compared to these techniques, BDCI is not limited to textual conflicts, which are usually easy to detect, but is designed to address higher-order conflicts, which are more challenging to detect~\cite{Brun:EarlyDetection:TSE:2013,Bang:WorkspaceAwarenessModeling:CSCW:2012,sarma:empirical:FSE2008}.

\balance
\textbf{Detection of Higher-Order Conflicts using Structural Information.} Higher-order conflicts are conflictual changes that do not cause textual conflicts but produce syntactic or semantic problems in the merged version of the program. These conflicts are particularly difficult to detect~\cite{Brun:EarlyDetection:TSE:2013,Bang:WorkspaceAwarenessModeling:CSCW:2012,sarma:empirical:FSE2008} and painful to fix~ \cite{brown:software:2002,Horwitz:NoninterferingVerions:TOPLAS:1989}.

Several techniques addressed the problem of detecting\linebreak higher-order conflicts by computing a representation of the dependencies between the components in a program, for instance using a program dependency graph or an abstract syntax tree. This representation is used to track the changes implemented by the developers and detect possible conflicts, such as concurrent changes on dependent artifacts.

The notable solutions implementing this approach differ on the representation of the program dependencies used to identify potential conflicts. CollabVS~\cite{dewan:semi-synchronous:ECSCW2007} uses a program call graph; TUKAN~\cite{schummer:TUKAN:ECSCW2001} uses a program def-use graph; Palantír~\cite{sarma:palantir:TSE2012} uses a program dependency graph; and Syde~\cite{Hattori:CollaborativeSWDevelopment:ICSE:2010} uses an abstract syntax tree.

Program dependencies are an important source of information for potential conflicts, but not all changes on dependent artifacts result in higher-order conflicts, and reporting any potential conflict, including the ones originated by indirect dependencies, might overwhelm users of false alarms. While reporting only changes to directly dependent artifacts may miss important and tricky conflicts. BDCI overcomes these issues by working on the behavior of the program, rather than on program dependencies. In this way, BDCI can exactly establish how and where a change impacts on the behavior of the program and precisely determine if two concurrent changes, even in remotely dependent code fragments, may cause interferences once merged.

\textbf{Detection of Higher-Order Conflicts using Behavioral Information.} Reporting the potential conflicts by looking at the program dependencies still requires developers to analyze the program changes in details to determine the presence of actual conflicts. As done in BDCI, a few other techniques tried to improve the detection of higher-order conflicts by looking at the impact of changes.

In particular, Safe-commit~\cite{wloka:safe-commit:2009} runs test cases in the background to proactively identify the changes that can be safely committed; Crystal~\cite{brun:crystal:ESECFSE2011,Brun:EarlyDetection:TSE:2013} and WeCode~\cite{Guimaraes:EarlyConflictDetection:ICSE:2012,guimaraes:towards:ICSE_CHASE2010} speculatively execute merge operations locally to the developer's workspace, and run the build and test processes, to discover interfering changes. Specific strategies can be adopted to reduce the cost of test executions~\cite{Nguyen:VariabilityAwareConflictDetection:ESECFSE:2015}. These techniques share with BDCI the idea to look at the behavior of the program to reveal higher-order conflicts. However, they only check the portion of the program behavior that is already checked by the oracles in the test cases (e.g., assert statements). Thus, they can discover a conflict only if it causes a test failure.

BDCI further elaborates the idea of working at the behavioral level to detect conflicts by explicitly generating a representation of the program behavior and studying the impact of concurrent changes on this representation. While the strategy based on test case execution only checks the input-output behavior of a program, the strategy based on behavioral models implemented in BDCI can detect interfering changes independently on the output generated by a program and the oracles in the test cases. The empirical results reported in this paper provide evidence of the complementarity between these approaches.

Finally, when the tasks that developers must execute are known a-priori and tracked by the development infrastructure, the Cassandra approach can be used to recommend developers task orders that minimize the chance to introduce conflicts~\cite{kasi:cassandra:ICSE2013}. BDCI targets the frequent case this knowledge is not available, and supports the detection of the conflicts that cannot be prevented by looking at task ordering.

\textbf{Detection of Higher-Order Conflicts in Non-Code\linebreak Artifacts.} Recently, the detection of higher-order conflicts gained attention also in the context of collaborative software design, where higher-order conflicts target design models rather than source code.\linebreak FLAME~\cite{bang:proactive:WICSA:2015,Bang:DesignConflicts:PHDThesis:2015} is a recently defined technique for proactive detection of higher-order software design conflicts. Although targeting a different context, it would be interesting to consider the possibility to extend FLAME with a conflict detection analysis working on the model semantics, similarly to what BDCI does for the code.

%% file: conclusions.tex
\section{Conclusions}\label{sec:conclusions}

Multi-branch development of software applications might be challenging and requires proper methods and tools to be performed efficiently. Modern SCM systems, such as 
Git~\cite{GIT:WEBSITE}, encourages the use of the branching logic, but also increases the number of conflicts that might be introduced while working on multiple concurrent branches.

While the detection and resolution of textual conflicts is extensively supported, the detection and resolution of \emph{higher-order conflicts} is still painful and only partially supported. In this paper we presented Behavioral Driven Conflict Identification (BDCI), an approach that introduces the novel idea to \emph{raise conflict detection from the source code level to the behavioral level}. The approach leverages two key technologies: \emph{specification mining}, to automatically generate models that represent the behavior of the software, and \emph{model analysis}, to identify the behaviors modified in each branch and the changes that might result in misbehaviors once merged.

BDCI is a general approach that can be potentially applied to multiple aspects of the software under analysis. In this paper we presented a \emph{specific instance of BDCI}, namely BDCI$_f$, that shows how this analysis can be used to reveal higher-order conflicts that may impact on the functional behavior of an application. 

%% file: artifact.tex
\section{Artifact Description}

Our artifact includes the data necessary to replicate the results obtained with BDCI$_f$ for Redis and Git.
The artifact is available for download at the following URL: \url{https://drive.google.com/drive/folders/0B8J_pv7c6buveTUxQkdnOXBoclk}.
Up to date information about the BDCI toolset can be found at \url{http://www.lta.disco.unimib.it/tools/bdci/}.

The artifact consists of a Virtualbox\footnote{www.virtualbox.com} virtual machine with installed Ubuntu, BDCI including its dependencies, a set of scripts for re-executing BDCI on each subject, and the rest of the data necessary to replicate the results reported in this paper. 
In particular, the virtual machine includes the source code, the test cases, and the intermediate results generated by BDCI$_f$ (e.g., the data recorded while running the test cases, and the models inferred with Daikon) for each set of versions that has been analyzed. The intermediate results can be used to execute the model analysis phase without having to execute the test cases and generate the models.


In general, our artifact can be exploited for multiple purposes:
\begin{itemize}
\item It enables other researchers to replicate the results reported in the paper.
\item It provides a significant amount of data collected during the execution of the test cases available with Git and Redis. Such data can be used for different research purposes by other researchers.
\item It provides a set of scripts that can be adapted to run BDCI on different program versions.
\item It provides a set of changes that introduce behavioral conflicts. These changes can be reused by other researchers to evaluate their techniques.
\end{itemize}

\subsection{Structure of the Artifact}

The artifact is organized as a Virtualbox virtual machine with four virtual disks: one disk contains BDCI$_f$ and the operating system, one disk can be used as working directory to run the analysis, and the other two disks contain the data for Git and Redis.
The virtual machine is large, it occupies ~100Gb of disk space after download. 

%
%
%
%


The data and the results generated for each case considered in our experiments have been compressed in a tar.gz archive and needs to be decompressed before running the analysis.

\subsection{Replicability}

Each case corresponds to a folder in the artifact. Each folder contains the script \texttt{replicate.sh} that can be executed to run the analysis. The output of the analysis is saved in the file \texttt{bdci.out}.
Note that each folder already contains the pre-computed \texttt{bdci.out} file. The file \texttt{bdci.out} is overwritten when \texttt{replicate.sh} is executed.

In the following we succinctly report the results generated by BDCI$_f$. For each case, we report conflicts using the syntax 

\begin{center} \texttt{Model} x \texttt{<->} y: (model for x) \texttt{<->} (model for y) \end{center}

\noindent where x and y could be 0, 1, or 2, where 0 indicates the base version of the software, and 1 and 2 indicate the two versions developed in parallel. The part \texttt{Model} x \texttt{<->} y identifies the two versions with incompatible models, while the part after : reports the models. We use the identifiers as shown in the dataset to identify cases.

%
%
%
%

\subsubsection{Git cases 171, 173, 181, 194, 197}

These cases produce higher-order conflicts that affect the functionality that prints the content of the packets processed by Git to the standard output. Figure~\ref{fig:artifact:171} shows the output generated by BDCI$_f$ for these five cases. The changes in the two branches affect the heading printed by the function \texttt{packet\_trace}. 
BDCI$_f$ detects this interference from the pre-condition of function \texttt{trace\_strbuf\_fl}, which is used by function \texttt{packet\_trace}.

\begin{figure}[htp]
\begin{center}
\footnotesize
\begin{tabular}{|p{8cm}|}
HIGHER-ORDER CONFLICT: function $trace\_strbuf\_fl\_$ENTER\\
Model 0<->1: (and (and (and (not ( = data 0)) (not ( = file 0)) ) (not ( = key 0)) ) (= line 46) )<->(and (and (and (not ( = data 0)) (not ( = file 0)) ) (not ( = key 0)) ) (= line 80) );\\
Model 0<->2:
(and (and (and (not ( = data 0)) (not ( = file 0)) ) (not ( = key 0)) ) (= line 46) )<->(and (and (and (not ( = data 0)) (not ( = file 0)) ) (not ( = key 0)) ) (= line 74) );\\
Model 1<->2:
(and (and (and (not ( = data 0)) (not ( = file 0)) ) (not ( = key 0)) ) (= line 80) )<->(and (and (and (not ( = data 0)) (not ( = file 0)) ) (not ( = key 0)) ) (= line 74) );\\
\end{tabular}
\end{center}\vspace{-0.5cm}
\caption{Results for Git cases 171, 173, 181, 194, 197.}
\label{fig:artifact:171}
\end{figure}%

\subsubsection{Git case 1357}

Figure~\ref{fig:artifact:1357} shows the output generated by BDCI$_f$.
This case is charachterized by a higher-order conflict that affects function \texttt{get\_wcwidth}, which computes the length of a Unicode character. 
BDCI$_f$ detects this problem from the pre-condition of function \texttt{bisearch}, which is used in function \texttt{get\_wcwidth}.

\begin{figure}[htp]
\begin{center}
\footnotesize
\begin{tabular}{|p{8cm}|}
HIGHER-ORDER CONFLICT: function bisearch$\_$ENTER\\
Model 0<->1: (and (= max 122) (not ( = table 0)) )<->(and (or (= max 51) (= max 231)) (not ( = table 0)) );\\
Model 0<->2: (and (= max 122) (not ( = table 0)) )<->(and (= max 119) (not ( = table 0)) );\\
Model 1<->2: (and (or (= max 51) (= max 231)) (not ( = table 0)) )<->(and (= max 119) (not ( = table 0)) );\\
\end{tabular}
\end{center}\vspace{-0.5cm}
\caption{Results for the Git case 1357.}
\label{fig:artifact:1357}
\end{figure}%

\subsubsection{Git case 131}

Figure~\ref{fig:artifact:131} shows the output generated by BDCI$_f$. The functions \texttt{cmd\_clone} and \texttt{checkout} can return different values in different branches.
This behaviour depends on the changes in function \texttt{wait\_or\_whine}. This may create problems in callers that process the returned values.

\begin{figure}[htp]
\begin{center}
\footnotesize
\begin{tabular}{|p{8cm}|}
HIGHER-ORDER CONFLICT: function cmd\_clone\_EXIT\\
Model 0<->1: (= return 0)<->(or (= return 0) (= return 128));\\
Model 0<->2:(= return 0)<->(or (= return 0) (= return 1));\\
Model 1<->2: (or (= return 0) (= return 128))<->(or (= return 0) (= return 1));\\
\end{tabular}
\end{center}\vspace{-0.5cm}
\caption{Results for the Git case 131.}
\label{fig:artifact:131}
\end{figure}%

\subsubsection{Redis case 76}

Figure~\ref{fig:artifact:1357} shows the output generated by BDCI$_f$.
The conflict targets function \texttt{rdbCheckThenExit}, which is used to trace errors in the Redis database management system.
A change in the caller code makes the function log a different line number in case of errors (see the values associated with parameter \texttt{where}). This could create issues in tools that process log files.

\begin{figure}[htp]
\begin{center}
\footnotesize
\begin{tabular}{|p{8cm}|}
HIGHER-ORDER CONFLICT: function rdbCheckThenExit\_ENTER\\
Model 0<->1:
(= where 1385)<->(= where 1498);\\
Model 0<->2:
(= where 1385)<->(= where 1400);\\
Model 1<->2:
(= where 1498)<->(= where 1400);\\
\end{tabular}
\end{center}\vspace{-0.5cm}
\caption{Results for the Redis case 76.}
\label{fig:artifact:76}
\end{figure}%